\documentclass[pra,aps,twocolumn,showpacs,floatfix,superscriptaddress]{revtex4-1}

\usepackage{amssymb,amsmath,amsfonts,graphicx,color,psfrag,times}

\newcommand{\nK}{\ensuremath{\mathrm{nK}}}
\newcommand{\um}{\ensuremath{\mathrm{\mu m}}}

\newcommand{\rbf}{\ensuremath{\mathbf{r}}}
\newcommand{\pbf}{\ensuremath{\mathbf{p}}}

\newcommand{\kB}{\ensuremath{k_{\mathrm{B}}}}
\newcommand{\Tc}{\ensuremath{T_{\mathrm{c}}}}

\begin{document}

\title{Growth dynamics of a Bose-Einstein condensate in a dimple trap without cooling}
\author{Michael~C.~Garrett}
\email{mgarrett@physics.uq.edu.au}
\affiliation{The University of Queensland, School of Mathematics and Physics, ARC Centre of Excellence for Quantum-Atom Optics, QLD 4072, Australia}
\author{Adrian~Ratnapala}
\altaffiliation[Current address: ]{Centre for Cold Matter, Blackett Laboratory, Imperial College London, London SW7 2AZ, United Kingdom}
\affiliation{The University of Queensland, School of Mathematics and Physics, QLD 4072, Australia}
\author{Eikbert~D.~van~Ooijen}
\affiliation{The University of Queensland, School of Mathematics and Physics, QLD 4072, Australia}
\author{Christopher~J.~Vale}
\altaffiliation[Current address: ]{ARC Centre of Excellence for Quantum-Atom Optics and Centre for Atom Optics and Ultrafast Spectroscopy, Swinburne University of Technology, Melbourne, VIC 3122, Australia}
\affiliation{The University of Queensland, School of Mathematics and Physics, QLD 4072, Australia}
\author{Kristian~Weegink}
\affiliation{The University of Queensland, School of Mathematics and Physics, QLD 4072, Australia}
\author{Sebastian~K.~Schnelle}
\affiliation{The University of Queensland, School of Mathematics and Physics, QLD 4072, Australia}
\author{Otto~Vainio}
\altaffiliation[Current address: ]{Turku Centre for Quantum Physics, Department of Physics and Astronomy, University of Turku, FI-20014 Turku, Finland}
\affiliation{The University of Queensland, School of Mathematics and Physics, QLD 4072, Australia}
\author{Norman~R.~Heckenberg}
\affiliation{The University of Queensland, School of Mathematics and Physics, QLD 4072, Australia}
\author{Halina~Rubinsztein-Dunlop}
\affiliation{The University of Queensland, School of Mathematics and Physics, QLD 4072, Australia}
\author{Matthew~J.~Davis}
\affiliation{The University of Queensland, School of Mathematics and Physics, ARC Centre of Excellence for Quantum-Atom Optics, QLD 4072, Australia}


\begin{abstract}

We study the formation of a Bose-Einstein condensate in a cigar-shaped three-dimensional harmonic trap, induced by the controlled addition of an attractive ``dimple''  potential along the weak axis. In this manner we are able to induce condensation without cooling due to a localized increase in the phase space density.  We perform a quantitative analysis of the thermodynamic transformation in both the sudden and adiabatic regimes for a range of dimple widths and depths.  We find good agreement with equilibrium calculations based on self-consistent semiclassical Hartree-Fock theory describing the condensate and thermal cloud.  We observe there is an optimal dimple depth that results in a maximum in the condensate fraction.  We also study the non-equilibrium dynamics of condensate formation in the sudden turn-on regime, finding good agreement for the observed time dependence of the condensate fraction with calculations based on quantum kinetic theory. 

\end{abstract}

\maketitle

\section{Introduction}

The formation of a Bose-Einstein condensate (BEC) and the growth of long-range coherence from a gas of thermal atoms is a problem of interest in the field of ultracold atoms~\cite{Stoof2007a}.  Before the first observations of Bose-Einstein condensation in a dilute gas there was some disagreement about the expected time scale for condensate formation~\cite{Griffin1995}. The first quantitative predictions  were made by Gardiner \emph{et al.}~\cite{Gardiner1997b}, who derived a rate equation for the growth of a single condensate mode from a super-critical thermal vapor. This was soon followed by the first experimental measurements of condensate formation by Miesner \emph{et al.}~\cite{Miesner1998b}.  Starting from just above the critical temperature for a BEC, this experiment implemented a sudden evaporative cooling ramp to remove the high-energy tail of a near-degenerate Bose gas. The ensuing rethermalization led to the formation of a Bose-Einstein condensate.  These experiments were subsequently analyzed using improved formalisms by Gardiner and co-workers~\cite{Gardiner1998b,Lee2000a,Davis2000a} and Bijlsma \emph{et al.}~\cite{Bijlsma2000c}.  They found that, while their numerical calculations were qualitatively in agreement with experimental observations, quantitatively no agreement could be found, and this has remained unresolved.

In 2002  K\"{o}hl \emph{et al.}~\cite{Kohl2002a} performed an experiment similar to that of  Miesner \emph{et al.}~\cite{Miesner1998b} but implemented continuous rather than sudden evaporation from near quantum degeneracy.  For this experiment the data were generally in good agreement with the results of quantum kinetic calculations incorporating the details of the evaporation and the effects of three-body loss~\cite{Davis2002a}.  The same formalism was applied to later  experiments in a quasi-condensate geometry by Hugbart~\emph{et al.}~\cite{Hugbart2007a}, where the calculated shape of the condensate growth curves agreed well with experiment apart from an unexplained time delay.  Other evaporative cooling experiments leading to BECs worth noting are those of Schvarchuck \emph{et al.}~\cite{Shvarchuck2002a}, who performed shock cooling in an elongated geometry and observed nonequilibrium dynamics in the resulting quasicondensate, and Ritter \emph{et al.}~\cite{Ritter2007a}, who measured the dynamics of the onset of long-range coherence in a three-dimensional condensate. 

Bose-Einstein condensation \emph{without} evaporative cooling was first induced by Stamper-Kurn \textit{et al.}~\cite{Stamper-Kurn1998c}, motivated by the earlier work of Pinkse \emph{et al.}~\cite{Pinkse1997a}.   Stamper-Kurn \textit{et al.} began with a near-degenerate Bose gas in a cigar-shaped harmonic trap and slowly applied an additional attractive ``dimple'' trap formed by a red-detuned optical dipole potential to adiabatically increase the phase-space density  by a factor of 50. It was shown that this was reversible within the limits of heating caused by their dipole trap.  Condensation was also induced by distillation without cooling in a double-well potential, demonstrated in an experiment by Shin \emph{et al.}~\cite{Shin2004a}. The lowering of a second well in this system caused the condensate in the first well to evaporate and re-form in the second at a higher temperature.   Erhard~\emph{et al.} observed the formation of an $m_F = 0$ BEC through spin collisions in an $F=1$ spinor condensate from initial partically condensed components in the $m_F=\pm1$ states, and they modeled their experiment using rate equations~\cite{Erhard2004a}. Recently an alternate approach to reversible BEC formation was demonstrated in an experiment by Catani \emph{et al.}~\cite{Catani2009a}, where entropy was exchanged between two atomic species, instead of between atoms inside and outside a dimple potential. 

In this paper we revisit the method of Stamper-Kurn \textit{et al.}~\cite{Stamper-Kurn1998c} to quantitatively study the dynamics and the thermodynamics of Bose-Einstein condensation, and we make comparisons of our experimental results with theoretical calculations.  We induce condensate formation by the controlled application of a tightly focused laser sheet to a near-degenerate Bose gas in a cigar-shaped magnetic trap (illustrated in Fig.~\ref{fig:dimple}). The addition of the resulting one-dimensional dimple potential to the weakly confined dimension of the harmonic trap induces condensation by locally increasing density while the temperature remains almost constant, hence increasing the local phase-space density.
 
We divide our results into two sections. First, we have measured the final equilibrium state of the Bose gas following both the quasistatic (i.e., slow) and sudden turn-on of the dimple potential for a range of laser intensities beginning from a well-controlled initial nondegenerate state.   The thermodynamics for an ideal gas with a delta function dimple have previously been studied in Ref.~\cite{Uncu2007a}.
Using a self-consistent mean-field model involving semiclassical Hartree-Fock theory for the thermal cloud and the Thomas-Fermi approximation for the condensate, we can predict the final condensate fraction for a given dimple depth for both the quasistatic and sudden turn-on.  We perform a quantitative comparison of experiment and theory for a thermodynamic transformation  through the BEC phase transition in an interacting Bose gas.
Second, we observe the dynamics of condensate formation following sudden turn-on of the dimple potential and compare with a quantum kinetic model of condensate growth.  This configuration allows a quantitative comparison with theory for condensate formation without evaporative cooling.  We note that this scenario has been studied previously using stochastic classical fields in one dimension~\cite{Stoof2001A,Proukakis2006a}.

This paper is organized as follows: In Sec.~\ref{sec:exp} we summarize our experimental setup and procedure. In Sec.~\ref{sec:thermo} we present our study of equilibrium thermodynamics by comparing the theoretical predictions with our experimental data and discussing the results. In Sec.~\ref{sec:dyn}, we present our study of condensate formation dynamics, first providing details of our quantum kinetic theory, and then comparing the theoretical predictions with our experimental data and discussing results. We finish with conclusions in Sec.~\ref{sec:conc}, and we provide additional theoretical details in the appendices.

\section{Experimental procedure}
\label{sec:exp}

Our experiments are performed on a gas of ultracold $^{87}$Rb atoms in the $5^2S_{1/2}|F=1,m_F=-1\rangle$ state, confined in a magnetic harmonic potential provided by a Z-wire configuration on an atom chip, as previously described in Ref.~\cite{Vale2004a}.  Briefly, we collect $\sim 10^7$ atoms in a reflection MOT before transferring them to a magnetic trap at a distance of 200~$\mu$m below the chip surface with axial and transverse frequencies of our cylindrically symmetric trapping potential $(\omega_z, \omega_\perp) \approx 2\pi \times(20, 1600)$~Hz.  Using rf evaporation we cool to near quantum degeneracy before decreasing the bias magnetic field to move the trap to 430~$\mu$m below the chip surface, with a resulting decrease in trapping frequencies to $(\omega_z, \omega_\perp) \approx 2\pi \times(6.8, 160)$~Hz such that the resulting system is more three dimensional and less susceptible to phase fluctuations that exist in elongated condensates in the tighter trap~\cite{Petrov2001a}. Further evaporative cooling results in a cloud of $(\sim 2$ -- $6)\times 10^5$ atoms. We go as close as we can to the BEC transition while remaining above the transition temperature. This ensures the largest possible condensate fraction when the dimple potential is subsequently turned on.

\begin{figure}
  \begin{center}
  \includegraphics[width=0.44\textwidth]{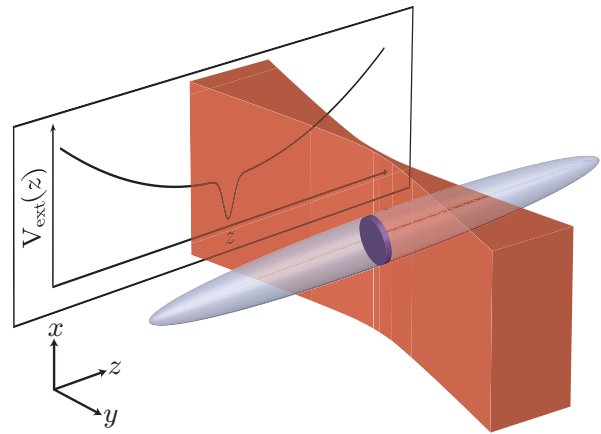}
  \end{center}
\caption{(Color online) Schematic of the experiment.  We create a dimple potential by tightly focusing a laser sheet at the center of the cigar-shaped harmonic magnetic potential. The total external potential $V_{\mathrm{ext}}(z)$ along the axial dimension then consists of a wide harmonic potential plus a narrow Gaussian dimple potential, while the potential in the other two spatial dimensions is unaffected.}
\label{fig:dimple}
\end{figure}

Starting from these initial conditions, we next apply a red-detuned $\lambda = 840$ nm optical dipole potential to the system, intersecting the weak ($z$) direction of the magnetic trap at the center as illustrated in Fig.~\ref{fig:dimple}. This dipole beam is known by direct charge-coupled device (CCD) imaging to be Gaussian and diffraction limited. By observing the effect of shifting the focus back and forth along the $y$ dimension, we ensured the magnetic trap intersects the beam waist. The beam can be focused to two different $1/e^2$ half-widths in the $z$ dimension, thereby creating either a wide (32 $\mu$m) or a narrow (11 $\mu$m) dimple potential. This should be compared to the typical thermal cloud extent of $\sim 400$ $\mu$m near the critical temperature. A cylindrical lens is used to expand the beam to widths of 350~$\mu$m (wide) or 220~$\mu$m (narrow) in the perpendicular dimension compared to the cloud width of 6~$\mu$m, such that the intensity is approximately constant in these dimensions. Thus the cylindrically symmetric trapping potential can be approximated by
\begin{eqnarray} 
V_{\mathrm{ext}}(r,z,t) &=& \frac12 m(\omega_\perp^2 r^2 + \omega_z^2 z^2) - A(t) e^{-2(z/w)^2},
\end{eqnarray} 
where $w$ is the $1/e^2$ half-width and the optical potential depth
\begin{equation}
A(t) = \frac{1}{2} I_0(t) \sum_{k\in \{D_1,D_2\}} \frac{\sigma_k \gamma_k}{\omega_k (\omega_k - \omega_{\mathrm{L}})},
\end{equation}
is proportional to the peak laser intensity $I_0(t)$, and can reach a maximum depth of $A(t)/\kB = 210$ nK (wide) or 1610 nK (narrow). The other relevant parameters are the scattering cross sections $\sigma_k$, linewidths $\gamma_k$, the resonant frequencies $\omega_k$ of the $D_1$ and $D_2$ lines in $^{87}$Rb, and the laser frequency $\omega_{\mathrm{L}}$ \cite{Grimm2000a}.

Our measurements are performed using absorption imaging after a time of flight of 20.3 ms, after turning off all trapping potentials. The temperature is determined by fitting a thermal cloud distribution to the wings of the resulting image, and the condensate fraction is determined from a two-component fit to the density. The condensate is sufficiently dense that condensate fractions of less that 1\% can be distinguished from the thermal background.

\section{Thermodynamic transformations across the critical point}
\label{sec:thermo}

In our first set of experiments, we apply dimples of various depths and measure the resulting final equilibrium temperatures and condensate fractions. We do this for both wide and narrow dimples, and for both quasistatic and sudden turn-on. We then compare our measured values with the predictions of semiclassical Hartree-Fock theory, incorporating full mean-field interactions of both the condensate and thermal cloud, as well as accounting for the effects of three-body loss. While it is widely assumed that this is appropriate for the three-dimensional Bose gas, there have been relatively few comprehensive comparisons with experimental data aside from Ref.~\cite{Gerbier2004b}. 

\subsection{Theoretical procedure}

We determine the initial entropy $S_\mathrm{i}$ and total energy $E_\mathrm{i}$ prior to dimple turn-on [$A(t=0)=0$], given the experimentally measured initial temperature and atom number, using the semiclassical theory as outlined in Appendix~\ref{sec:sc}. To predict the final state, 
in the case of quasistatic turn-on we assume the system evolves isentropically and solve for the final temperature at which $S=S_\mathrm{i}$. In the case of sudden turn-on, we use the initial density to calculate the sudden change in energy imparted by the dimple,
\begin{equation}
 \Delta E = -\int d\rbf Ae^{-2(z/w)^2} [n_0(\rbf) + n_\mathrm{th}(\rbf)] \mathrm{,}
\end{equation}
where $n_0(\rbf)$ and $n_\mathrm{th}(\rbf)$ are the densities of the condensate and thermal cloud, respectively. We then solve for the final temperature at which $E=E_\mathrm{i}+\Delta E$, which determines the final condensate fraction. We also estimate the effects of three-body loss in our calculations, as detailed in Appendix~\ref{sec:3B}.

\subsection{Comparison with experiment --- wide dimple}

Our experimental procedure is as follows.  For the wide dimple measurements, we begin with an atomic cloud of $N=6.25(25)\times 10^5$ atoms at a temperature of $T_\mathrm{i}=215(2)$~nK. This corresponds to a phase-space density at the center of the trap of $\Gamma\approx 2.6$, indicating that the cloud is very close to the BEC transition ($\Gamma=\zeta(3/2)\approx 2.612$). To turn on the dimple quasistatically, we ramp up the dimple potential linearly at a rate of 70$\kB~\nK/$s, and then hold it constant for a 300-ms equilibration time before turning off all potentials and imaging. To ensure that we are in the quasistatic regime, we have repeated this process for various ramp rates: For faster rates we observe a decrease in condensate fraction at large dimple depths due to nonadiabatic heating.  To turn on the dimple suddenly, we ramp up the potential in less than 0.1~ms, and then hold it constant for a 1000-ms equilibration time before imaging. 

\begin{figure}
  \begin{center}
  \includegraphics[width=0.47\textwidth]{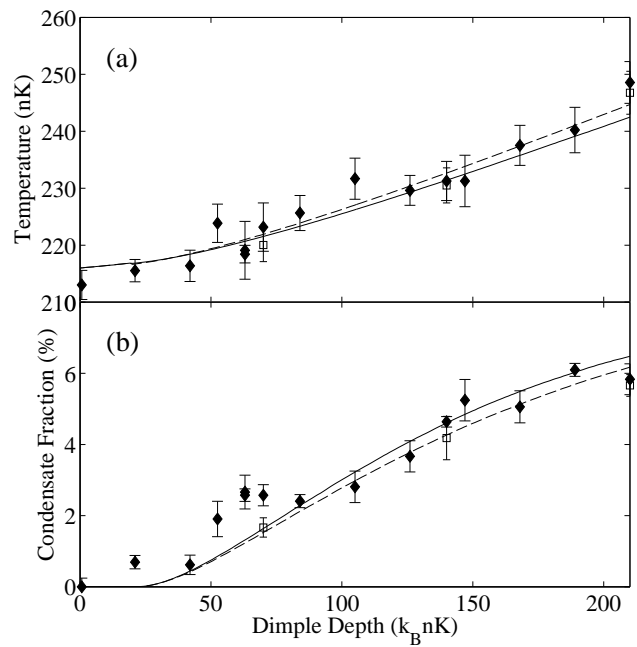}
  \end{center}
\caption{Comparison of theory and experiment for (a) the final temperature and (b) the condensate fraction at final equilibrium after the turn-on of the wide dimple potential. Quasistatic turn-on is indicated by solid lines for theory and by diamonds for experimental data.  Sudden turn-on is indicated by dashed lines for theory and by squares for experimental data. All experimental data points are four-measurement averages with the error bars indicating the standard deviation of the mean.}
\label{fig:wide}
\end{figure}

The resulting temperatures and condensate fractions are plotted versus final dimple depth and compared with the predictions of semiclassical theory in Fig.~\ref{fig:wide}.  We find good agreement between theory and experiment for both the quasistatic and sudden turn-on, with the best fits obtained using initial conditions $N=6.50\times 10^5$ and $T_\mathrm{i}=216$~nK. We also find the results of the quasistatic and sudden turn-on to be nearly indistinguishable from each other. This is because only a small fraction of the atoms are drawn from the harmonic trap into the wide dimple, which has a maximum attainable depth of order $\sim \kB T$.

We compare the predictions of the semiclassical theory with our experimental data for both wide (Fig.~\ref{fig:wide}) and narrow (Fig.~\ref{fig:narrow}) dimples. We plot temperature and condensate fraction versus dimple depth for both quasistatic and sudden turn-on. In all cases, the only fitting parameters used to generate the theoretical curves are the initial temperature and atom number, constrained to lie within their respective measurement uncertainties.

\subsection{Comparison with experiment --- narrow dimple}

Using the more tightly focused narrow dimple, we are able to attain a maximum depth much greater than $\kB T_\mathrm{i}$ and thereby observe differences between quasistatic and sudden turn-on. For the quasistatic turn-on we begin with an atomic cloud of $N=2.60(15)\times 10^5$ atoms at a temperature of $T_\mathrm{i}=160(2)$~nK ($\Gamma\approx 2.6$).  We ramp up the dimple potential linearly over a time of 1500~ms, and then hold the potential constant for a 300-ms equilibration time before imaging. For the sudden turn-on we begin with an atomic cloud of $N=2.10(15)\times 10^5$ atoms at a temperature of $T_\mathrm{i}=168(2)$~nK ($\Gamma\approx 1.1$) and  follow the same procedure as for the wide dimple.

\begin{figure}
  \begin{center}
  \includegraphics[width=0.47\textwidth]{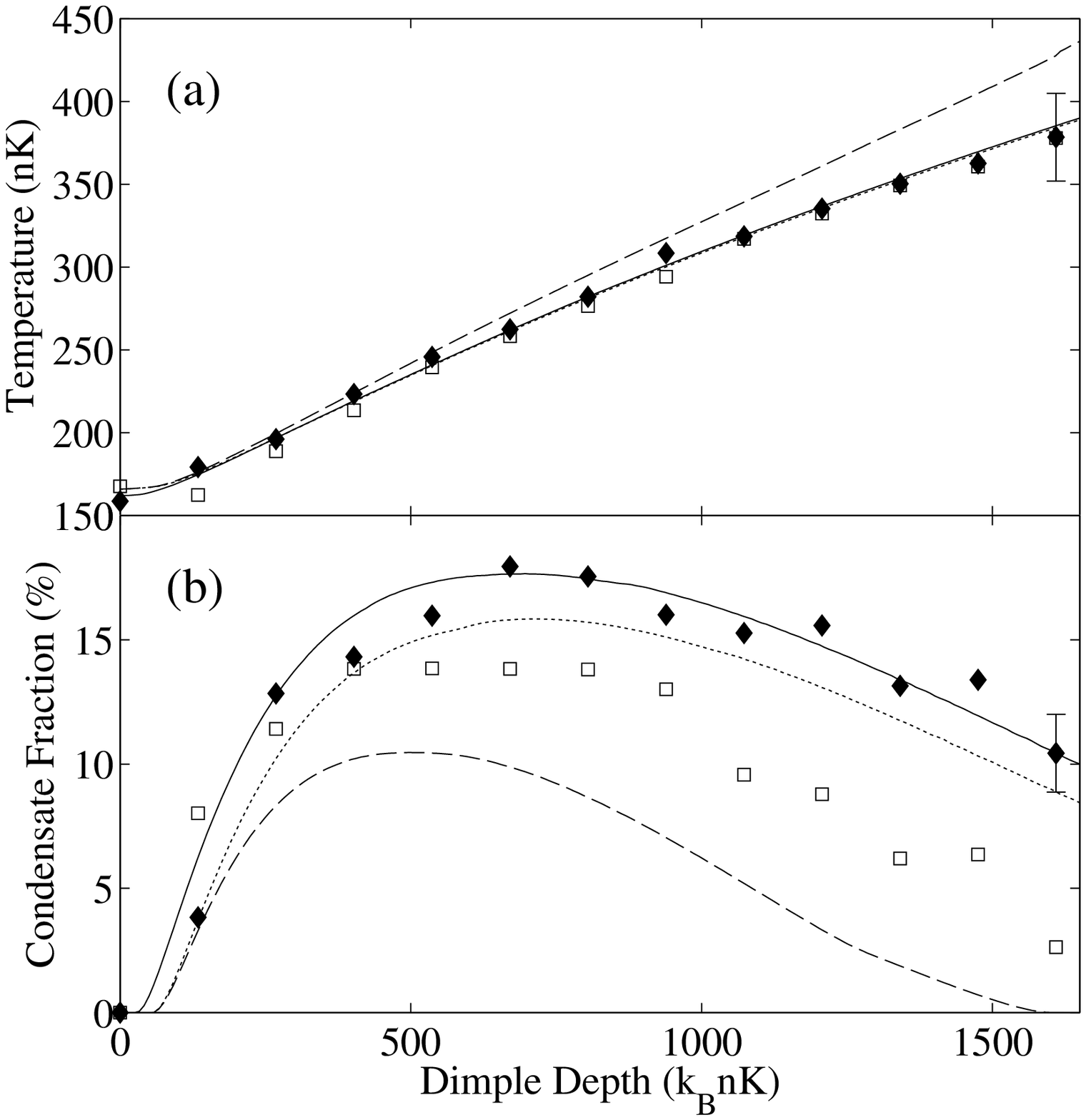}
  \end{center}
\caption{Comparison of theory and experiment for (a) the final temperature and (b) the condensate fraction at final equilibrium after the turn-on of the  narrow dimple potential. Quasistatic turn-on is indicated by solid lines for theory and by diamonds for experimental data.  Sudden turn-on is indicated by dashed lines for theory and by squares for experimental data. The dotted lines indicate quasistatic turn-on from the same initial conditions used for sudden turn-on. All experimental data points are single measurements, with the exception of a four-point measurement at depth $A/\kB=1610$~nK to measure the standard deviation of the mean.}
\label{fig:narrow}
\end{figure}

The resulting temperatures and condensate fractions are plotted versus final dimple depth and compared with theory in Fig.~\ref{fig:narrow}. In this case we observe a significant difference between the results of quasistatic and sudden turn-on, though this is partly due to the difference in initial conditions. We find good agreement between theory and experiment for the quasistatic turn-on, with the best fits obtained using initial conditions $N=2.45\times 10^5$ and $T_\mathrm{i}=162$~nK.

However, we do not find good agreement for the sudden turn-on, with theory predicting a much smaller condensate fraction than was experimentally observed. The best fit shown is for $N=2.25\times 10^5$ and $T_\mathrm{i}=166$~nK, though better fits can be obtained by using values of $N$ and $T_\mathrm{i}$ that lie outside their respective measurement uncertainties.

A potential explanation for this discrepancy is that the turn-on is not sufficiently quick, and the density of the gas is not frozen during the turn-on.  This would result in a smaller increase in total energy of the gas compared to the prediction of our model, leading to a smaller increase in temperature and hence a larger condensate fraction --- as is observed experimentally. Indeed, in the limit of quasistatic turn-on from the same initial conditions (dotted lines in Fig.~\ref{fig:narrow}), the predicted condensate fraction is considerably larger than for sudden turn-on. Since the measured values lie between the limits of quasistatic and sudden turn-on, an intermediate turn-on time would likely provide a good fit to the data.

To model this would require a fully dynamical treatment that is beyond the limitations of our semiclassical theory and is numerically impractical within the quantum kinetic model we introduce later in Sec.~\ref{sec:qk}. Furthermore, departures from the sudden case would only be expected at turn-on times similar to the time scale for rethermalization. Since the mean free time between collisions at initial conditions is approximately 6 ms, whereas the time for sudden turn-on is less than 0.1 ms, this explanation seems unlikely. We are therefore unfortunately forced to leave this discrepancy unresolved.

\subsection{Discussion}

\begin{figure}
  \begin{center}
  \includegraphics[width=0.47\textwidth]{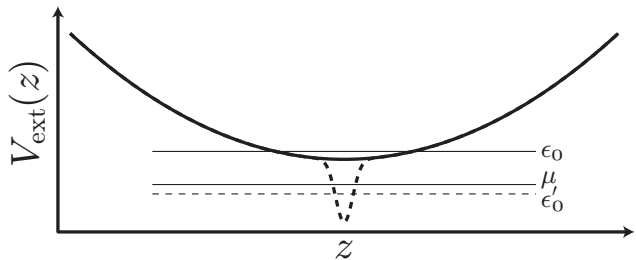}
  \end{center}
\caption{Schematic of the effect of dimple turn-on for small depths $A$. Before the dimple is turned on (solid lines), as the system temperature is above $\Tc$ the chemical potential $\mu$ lies below the translational ground state energy $\epsilon_0$ of the harmonic trap. When the dimple is turned on (dashed lines), the new ground state energy $\epsilon_0'$ lies below the initial chemical potential, inducing condensation. Due to interactions, the energy $\epsilon_0'$ increases as the condensate grows until equilibrium is reached ($\epsilon_0'\approx \mu$).}
\label{fig:epsilonmu}
\end{figure}

An interesting feature of the narrow dimple data is that there exists an optimal dimple depth ($\sim 750\kB$~$\nK$) at which a maximum condensate fraction is obtained. As the dimple depth is increased beyond this value, the condensate fraction gradually decreases back toward zero. We can understand this feature in the context of the condensate formation process as follows. At shallow depths, the dimple potential acts merely as a perturbation to the broader harmonic trap. To a first approximation the chemical potential and temperature are unchanged, whereas the energy of the translational ground state is decreased relative to the bottom of the harmonic trap (see Fig.~\ref{fig:epsilonmu}). When the ground-state energy approaches the chemical potential, a condensate forms, as observed for the wide dimple and at shallow depths of the narrow dimple. However, at larger depths a significant fraction of the thermal cloud is drawn into the narrow dimple, causing non-negligible changes in $\mu$ and $T$. At sufficiently large depths, the entire thermal cloud falls into the dimple potential (which is approximately harmonic near its center) and the condensate evaporates, in agreement with the well-known result that compression of a gas cannot alter the phase-space density~\cite{Houbiers1997a}.

\section{Condensate formation dynamics}
\label{sec:dyn}

In our second set of experiments, we suddenly turn on the dimple and allow the system to evolve for various times before measuring the condensate fraction. We do this for both wide and narrow dimples, in each case for two different fixed dimple depths: one shallow and the other deep. We then compare our measured values with the predictions of quantum kinetic theory. In particular, we use the ergodic quantum Boltzmann equation~\cite{Luiten1996a} with the additional inclusion of the effects of the condensate mean-field~\cite{Davis2000a,Bijlsma2000c} and three-body loss. While similar comparisons have been made previously~\cite{Bijlsma2000c,Davis2000a,Kohl2002a,Davis2002a,Hugbart2007a} the condensation transitions in these experiments were induced by evaporative cooling. Here condensation is introduced without loss by modifying the density of states of the trap.

\subsection{Quantum kinetic theory}
\label{sec:qk}

The initial equilibrium state of the Bose gas above $\Tc$ becomes nonequilibrium following the sudden turn-on of a dimple potential. We begin by calculating this initial nonequilibrium state, and then evolve it in time to final equilibrium using an ergodic quantum Boltzmann equation (EQBE), as previously described in~\cite{Luiten1996a,Davis2000a,Bijlsma2000c}.

The ergodic assumption is that all semiclassical states for a given single-particle energy $\epsilon$ have the same mean occupation and thus the nonequilibrium phase-space distribution function depends only on time and energy: $f(\rbf,\pbf,t)=f(\epsilon(\rbf,\pbf,t),t)$. Hence the position and momentum dependence in the full quantum Boltzmann equation is projected out, yielding a partial differential equation in energy and time only. The ergodic approximation is necessary in order to reduce the dimensionality of the phase space to render the quantum Boltzmann equation computationally tractable. The EQBE then governs the evolution of the energy density
\begin{equation}
\begin{split}
  n(\epsilon,t) & =\int \frac{d\rbf d\pbf}{(2\pi\hbar)^3} \delta\big(\epsilon-\epsilon_{\mathrm{th}}(\rbf,\pbf,t)\big) f(\rbf,\pbf,t) \\
  						  & =g(\epsilon,t)f(\epsilon,t) \mathrm{,}
\end{split}
\end{equation}
where $f(\epsilon,t)$ is the energy distribution function and
\begin{equation} \label{eq:dos}
g(\epsilon,t)=\int \frac{d\rbf d\pbf}{(2\pi\hbar)^3} 
  						\delta\big(\epsilon-\epsilon_{\mathrm{th}}(\rbf,\pbf,t)\big)
\end{equation}
is the density of states. The semiclassical excitation energy $\epsilon_{\mathrm{th}}(\rbf,\pbf,t)$ is as defined in Eq.~(\ref{eq:eth}) but with the mean field of the thermal cloud neglected [i.e.,~setting $n_{\mathrm{th}}(\rbf)$ to zero]. This eliminates the need to determine the densities self-consistently, and our results suggest that this approximation is reasonable. The condensate mean field is calculated via the Thomas-Fermi approximation, as per Eq.~(\ref{eq:TF}), but again neglecting the mean field of the thermal cloud. We provide a more detailed description of the EQBE in Appendix~\ref{sec:EQBE} and describe how to incorporate the effects of three-body loss in Appendix~\ref{sec:3B}.

\begin{figure}[b]
  \begin{center}
		\begin{psfrags}%
		\psfragscanon%
		%
		\psfrag{s05}[t][t]{\color[rgb]{0,0,0}\setlength{\tabcolsep}{0pt}\begin{tabular}{c}$\epsilon$ $(k_{\mathrm{B}}T_{\mathrm{i}})$\end{tabular}}%
		\psfrag{s06}[b][b]{\color[rgb]{0,0,0}\setlength{\tabcolsep}{0pt}\begin{tabular}{c}$n(\epsilon,t)$ (atoms$/\hbar\bar{\omega}$)\end{tabular}}%
		\psfrag{s10}[][]{\color[rgb]{0,0,0}\setlength{\tabcolsep}{0pt}\begin{tabular}{c} \end{tabular}}%
		\psfrag{s11}[][]{\color[rgb]{0,0,0}\setlength{\tabcolsep}{0pt}\begin{tabular}{c} \end{tabular}}%
		\psfrag{s12}[l][l]{\color[rgb]{0,0,0} $t=1000$ ms}%
		\psfrag{s13}[l][l]{\color[rgb]{0,0,0} $t<0$}%
		\psfrag{s14}[l][l]{\color[rgb]{0,0,0} $t=0$}%
		\psfrag{s15}[l][l]{\color[rgb]{0,0,0} $t=1000$ ms}%
		%
		\psfrag{x01}[t][t]{$-A$}%
		\psfrag{x02}[t][t]{0}%
		\psfrag{x03}[t][t]{5}%
		\psfrag{x04}[t][t]{10}%
		\psfrag{x05}[t][t]{15}%
		%
		\psfrag{v01}[r][r]{0}%
		\psfrag{v02}[r][r]{200}%
		\psfrag{v03}[r][r]{400}%
		\psfrag{v04}[r][r]{600}%
		\psfrag{v05}[r][r]{800}%
		%
		\includegraphics[width=0.47\textwidth]{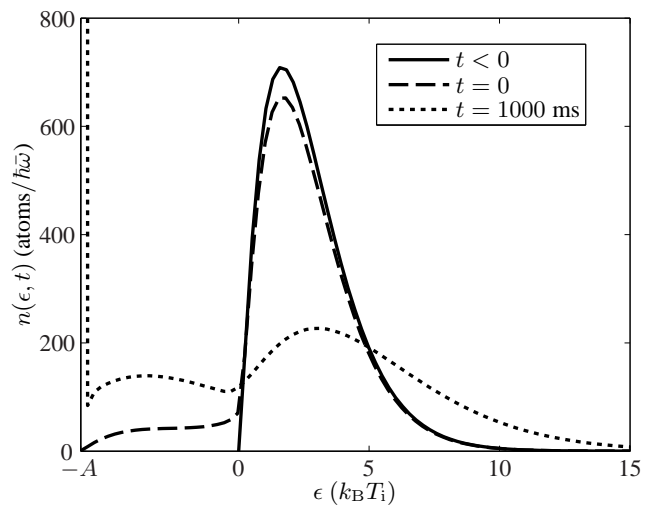}%
		\end{psfrags}%
  \end{center}
\caption{Energy density $n(\epsilon,t)=g(\epsilon,t)f(\epsilon,t)$ before dimple turn-on at $t<0$ (solid line), immediately after dimple turn-on at $t=0$ (dashed line), and at final equilibrium after $t=1000$~ms (dotted line). Strictly speaking, the value plotted at $\epsilon=-A$ is the condensate occupation, not the energy density. Immediately after dimple turn-on to depth $A$, only a small fraction of atoms have energies below the bottom of the harmonic potential ($\epsilon<0$). Over the 1000-ms equilibration time, this fraction steadily increases, eventually resulting in the formation of a condensate.}
\label{fig:edensity}
\end{figure}

A typical atom will travel a distance of less than 1~$\um$ during the sudden ramp-up of the dimple potential. As this is more than an order of magnitude smaller than the width of the narrow dimple, we approximate the sudden turn-on as instantaneous. Under this assumption, the phase-space distribution immediately after the sudden turn-on must be the same as before the turn-on: $f'(\rbf,\pbf)=f(\rbf,\pbf)=f_{\mathrm{BE}}\big(\epsilon_{\mathrm{th}}(\rbf,\pbf)\big)$, where we henceforth use primed (unprimed) variables to denote quantities immediately after (before) dimple turn-on at $t=0$. Although the phase-space distribution function is unchanged, the post-dimple semiclassical excitation energy differs from the pre-dimple expression via the inclusion of the dimple potential:
\begin{equation}
	\epsilon_{\mathrm{th}}'(\rbf,\pbf) = \epsilon_{\mathrm{th}}(\rbf,\pbf) - Ae^{-2(z/w)^2} \mathrm{.}
\end{equation}
Clearly, the change in $\epsilon_{\mathrm{th}}'(\rbf,\pbf)$ alters the density of states in Eq.~(\ref{eq:dos}), and therefore it alters the energy density, which we calculate immediately after the dimple turn-on as
\begin{equation}
	n'(\epsilon)=\int \frac{d\rbf d\pbf}{(2\pi\hbar)^3} \delta\big(\epsilon-\epsilon_{\mathrm{th}}'(\rbf,\pbf)\big)  
																												f_{\mathrm{BE}}\big(\epsilon_{\mathrm{th}}(\rbf,\pbf)\big) \mathrm{.}
\end{equation}
In doing this we are ergodically projecting the pre-dimple phase-space distribution using the post-dimple density of states. It should be noted that the physical initial phase space will actually be non-ergodic. However, previous Monte Carlo calculations of the Boltzmann equation have shown that ergodicity is restored relatively quickly~\cite{Wu1996b}, and hence we expect this should be a reasonable approximation.

\begin{figure}[t]
  \begin{center}
		\begin{psfrags}%
		\psfragscanon%
		%
		\psfrag{s07}[t][t]{\color[rgb]{0,0,0}\setlength{\tabcolsep}{0pt}\begin{tabular}{c}Time since dimple turn-on (ms)\end{tabular}}%
		\psfrag{s08}[b][b]{\color[rgb]{0,0,0}\setlength{\tabcolsep}{0pt}\begin{tabular}{c}Condensate fraction (\%)\end{tabular}}%
		%
		\psfrag{x01}[t][t]{0}%
		\psfrag{x02}[t][t]{50}%
		\psfrag{x03}[t][t]{100}%
		\psfrag{x04}[t][t]{0}%
		\psfrag{x05}[t][t]{200}%
		\psfrag{x06}[t][t]{400}%
		\psfrag{x07}[t][t]{600}%
		\psfrag{x08}[t][t]{800}%
		\psfrag{x09}[t][t]{1000}%
		%
		\psfrag{v01}[r][r]{0}%
		\psfrag{v02}[r][r]{1}%
		\psfrag{v03}[r][r]{2}%
		\psfrag{v04}[r][r]{3}%
		\psfrag{v05}[r][r]{0}%
		\psfrag{v06}[r][r]{1}%
		\psfrag{v07}[r][r]{2}%
		\psfrag{v08}[r][r]{3}%
		\psfrag{v09}[r][r]{4}%
		\psfrag{v10}[r][r]{5}%
		\psfrag{v11}[r][r]{6}%
		%
		\includegraphics[width=0.47\textwidth]{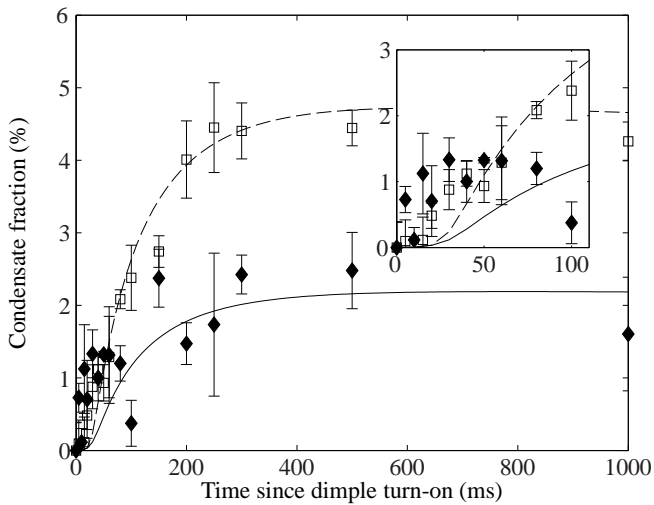}%
		\end{psfrags}%
		\vspace{-2 mm}
	\end{center}
\caption{Condensate growth curves for theory and experiment following the sudden turn-on of the wide dimple potential.  The condensate fraction is plotted vs time for dimple depths of $A/\kB = 70 $~nK (solid line: theory; diamonds: experiment) and $A/\kB = 140$~nK (dashed line: theory; squares: experiment). All experimental data points are four-measurement averages with the error bars indicating the standard deviation of the mean.}
\label{fig:widedyn}
\end{figure}

The pre- and post-dimple energy densities are shown in Fig.~\ref{fig:edensity}, for parameters typical of the narrow dimple: $A/\kB=805$~nK, $N=1.30\times 10^5$, and $T_\mathrm{i}=133$~nK. The post-dimple energy density closely resembles the pre-dimple energy density, except that the peak in post-dimple energy density is slightly smaller because a small fraction of atoms lying within the dimple now have energies below the bottom of the harmonic potential ($\epsilon < 0$). Also plotted in Fig.~\ref{fig:edensity} is the energy density at final equilibrium, calculated via the EQBE. As the system evolves toward equilibrium, successively more atoms are drawn into the dimple, occupying the energies in the range $-A<\epsilon<0$, and ultimately resulting in the formation of a condensate.

\begin{figure}[t]
  \begin{center}
		\begin{psfrags}%
		\psfragscanon%
		%
		\psfrag{s03}[t][t]{\color[rgb]{0,0,0}\setlength{\tabcolsep}{0pt}\begin{tabular}{c}Time since dimple turn-on (ms)\end{tabular}}%
		\psfrag{s04}[b][b]{\color[rgb]{0,0,0}\setlength{\tabcolsep}{0pt}\begin{tabular}{c}Condensate fraction (\%)\end{tabular}}%
		%
		\psfrag{x01}[t][t]{0}%
		\psfrag{x02}[t][t]{100}%
		\psfrag{x03}[t][t]{200}%
		\psfrag{x04}[t][t]{300}%
		\psfrag{x05}[t][t]{400}%
		\psfrag{x06}[t][t]{500}%
		\psfrag{x07}[t][t]{600}%
		\psfrag{x08}[t][t]{700}%
		\psfrag{x09}[t][t]{800}%
		\psfrag{x10}[t][t]{900}%
		\psfrag{x11}[t][t]{1000}%
		%
		\psfrag{v01}[r][r]{0}%
		\psfrag{v02}[r][r]{2}%
		\psfrag{v03}[r][r]{4}%
		\psfrag{v04}[r][r]{6}%
		\psfrag{v05}[r][r]{8}%
		\psfrag{v06}[r][r]{10}%
		\psfrag{v07}[r][r]{12}%
		\psfrag{v08}[r][r]{14}%
		%
		\includegraphics[width=0.47\textwidth]{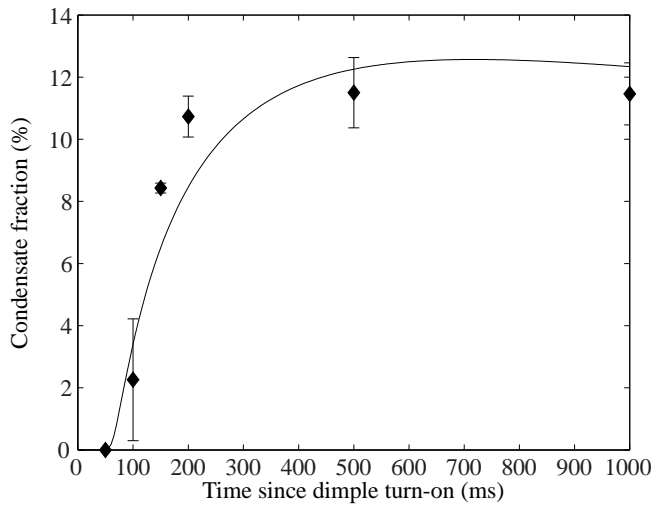}%
		\end{psfrags}%
		\vspace{-2 mm}
  \end{center}
\caption{Condensate growth curves for theory and experiment following the sudden turn-on of the narrow dimple potential.  The condensate fraction is plotted vs time for dimple depths of $A/\kB = 660 $ nK (solid line: theory; diamonds: experiment).  All experimental data points are four-measurement averages with the error bars indicating the standard deviation of the mean.}
\label{fig:narrowdyn}
\end{figure}

\subsection{Comparison with experiment}

We compare the predictions of the quantum kinetic calculations just described with our experimental data for both wide (Fig.~\ref{fig:widedyn}) and narrow (Fig.~\ref{fig:narrowdyn}) dimples. In both cases, we plot condensate fraction versus time for two different dimple depths. In all cases, the only fitting parameters used to generate the theoretical curves are the initial temperature and atom number, constrained to lie within their respective measurement uncertainties.

For the wide dimple measurements, we begin with an atomic cloud of $N=6.25(25)\times 10^5$ atoms at a temperature of $T_\mathrm{i}=215(2)$~nK ($\Gamma\approx 2.6$). We ramp up the potential suddenly, and then hold it constant for various equilibration times up to 1000~ms before imaging. We use dimples of two different depths: $A/\kB =70$ and $140$ nK. The resulting condensate fractions are plotted versus time and compared with the predictions of quantum kinetic theory in Fig.~\ref{fig:widedyn}. We find good agreement between theory and experiment, with the best fits obtained using initial conditions ($N=6.30\times 10^5$, $T_\mathrm{i}=217$ nK) and ($N=6.10\times 10^5$, $T_\mathrm{i}=216$ nK) for dimple depths $A/\kB =70$ and $140$ nK, respectively.

For the narrow dimple measurements, we begin with an atomic cloud of $N=1.87(7)\times 10^5$ atoms at a temperature of $T_\mathrm{i}=154(6)$~nK ($\Gamma\approx 1.4$), and we follow the same experimental procedure described earlier for a final dimple depth of $A/\kB =660$~nK. The results are plotted in Fig.~\ref{fig:narrowdyn}, with the best fit obtained using initial conditions $N=1.94\times 10^5$ and $T_\mathrm{i}=148$~nK. We find good agreement between theory and experiment, particularly at final equilibrium. In contrast to the wide dimple case, we observe a delay of about 50 ms in the onset of condensate formation, which agrees with the EQBE calculation.

\section{Conclusions}
\label{sec:conc}

We have studied the thermodynamics of the transition across the BEC critical point and the formation dynamics of Bose-Einstein condensation resulting from the application of attractive Gaussian dimple potentials of various widths and depths to a Bose gas in a cigar-shaped harmonic trap. We have measured the equilibrium temperature and condensate fraction over a range of dimple depths, for both quasistatic and sudden turn-on of the dimple, and compared our results with the predictions of semiclassical Hartree-Fock theory. For narrow dimples, we found that there exists an optimal dimple depth at which a maximum condensate fraction is attained. Beyond this depth, the dimple acts merely as a tighter harmonic trap, and it therefore does not increase the phase-space density. We found good agreement between theory and experiment, except in the case of sudden turn-on of deep, narrow dimples. We also measured the (nonequilibrium) condensate fraction over a range of times after sudden turn-on of the dimple and compared our results with the predictions of quantum kinetic theory. We observed a short delay in the onset of condensate formation in the case of narrow dimples, but not in the case of wide dimples. In both cases we found good agreement between theory and experiment. This provides further validation to the quantum kinetic model of condensate formation, for which previous comparisons were based on sudden evaporative cooling.

\begin{acknowledgments}

This research was supported under the Australian Research Council's Discovery Projects funding scheme (Project Nos. DP0343094 and DP0985142) and the ARC Centre of Excellence for Quantum-Atom Optics (CE0348178).  MCG acknowledges financial support from NSERC, Endeavour IPRS, and the University of Queensland.
\end{acknowledgments}

\appendix

\section{semiclassical theory}
\label{sec:sc}

We make use of the Thomas-Fermi approximation for the condensate density and the semiclassical Hartree-Fock approximation for the thermal cloud, as outlined in~\cite{Goldman1981a,Huse1982a,Oliva1989a,Bagnato1987a,Shi1997a,Shi1997b,Giorgini1997b} and compared with experiment by Gerbier \textit{et al.}~\cite{Gerbier2004b}. 

At equilibrium the thermal cloud is well described by the Bose-Einstein distribution, $f_{\mathrm{BE}}\big(\epsilon_{\mathrm{th}}(\rbf,\pbf)\big)=(\mathrm{exp}\{[\epsilon_\mathrm{th}(\rbf,\pbf)-\mu]/\kB T\}-1)^{-1}$, where $T$ is the temperature and $\mu$ is the chemical potential. In the semiclassical Hartree-Fock approximation, the excitation energy $\epsilon_\mathrm{th}(\rbf,\pbf)$ (the energy required to add an atom with momentum $\pbf$ at position $\rbf$) is given by the expression
\begin{equation}\label{eq:eth}
   \epsilon_\mathrm{th}(\rbf,\pbf)=\frac{p^2}{2m}+V_{\mathrm{ext}}(\rbf)+U_0\left[2n_0(\rbf)+2n_\mathrm{th}(\rbf)\right] \mathrm{,}
\end{equation}
where $m$ is the atomic mass and $U_0=4\pi\hbar^2 a/m$ is the interaction strength, proportional to the $s$-wave scattering length $a$. The mean-field density combined with the external potential $V_{\mathrm{ext}}(\rbf)$ constitute the effective potential experienced by atoms in the thermal cloud. The density of the thermal cloud is calculated as
\begin{equation}
   n_\mathrm{th}(\rbf)=\int \frac{d\pbf}{(2\pi\hbar)^3} f_{\mathrm{BE}}\big(\epsilon_{\mathrm{th}}(\rbf,\pbf)\big) \mathrm{,}
\end{equation}
while the density of the condensate is calculated using the Thomas-Fermi approximation,
\begin{equation}\label{eq:TF}
   n_0(\rbf)=\mathrm{max}\Big\{0,\big(\mu-\left[V_{\mathrm{ext}}(\rbf)+2U_0n_\mathrm{th}(\rbf)\right]\big)/U_0\Big\}\mathrm{.}
\end{equation}
Thus, the densities of both the thermal cloud and condensate must be solved self-consistently for a given $T$ and $\mu$ to give the experimentally measured total atom number $N=\int d\rbf \left[n_0(\rbf)+n_\mathrm{th}(\rbf)\right]$.

From here other thermodynamics quantities such as the condensate fraction $N_0/N$, total entropy, and total energy can be determined.  The last two are relevant to quasistatic and sudden turn-on of the dimple potential, respectively. By writing $\epsilon \equiv \epsilon_{\mathrm{th}}(\rbf,\pbf)$ the total entropy is given by (cf. p. 15 of \cite{Pitaevskii2003a})
\begin{eqnarray}
   S  &=&  \kB  \int  \frac{d\rbf d\pbf}{(2\pi\hbar)^3} 
   \bigg\{\Big[f_{\mathrm{BE}}(\epsilon)+1\Big]
   \mathrm{ln}\Big[f_{\mathrm{BE}}(\epsilon)+1\Big] \nonumber \\
	 & & \qquad\qquad\qquad - f_{\mathrm{BE}}(\epsilon)
	 \mathrm{ln}\Big[f_{\mathrm{BE}}(\epsilon)\Big]\bigg\} \mathrm{,}
\end{eqnarray}
while the total energy of the system is
\begin{eqnarray}
E &=& K + V + I \mathrm{,} \\
K &=& \int \frac{d\rbf d\pbf}{(2\pi\hbar)^3} \frac{p^2}{2m} f_{\mathrm{BE}}(\epsilon) \mathrm{,} \\
V &=& \int d\rbf V_{\mathrm{ext}}(\rbf) [n_0(\rbf) + n_\mathrm{th}(\rbf)] \mathrm{,} \\
I &=& \frac{U_0}{2}\int d\rbf \left[ n_0^2(\rbf) + 4 n_0(\rbf)n_\mathrm{th}(\rbf) + 2 n_\mathrm{th}^2(\rbf)\right] \mathrm{.}
\end{eqnarray}

\section{Three-body loss}
\label{sec:3B}

Three-body loss occurring between condensed and non-condensed atoms can be calculated from the three-body correlation function as
\begin{eqnarray}
\dot{n}(\rbf)_\mathrm{3B} = -K_3 \langle \hat{\Psi}^\dag(\rbf)^3\hat{\Psi}(\rbf)^3\rangle \mathrm{,}
\end{eqnarray}
where $K_3=5.8(1.9)\times 10^{-30}$~$\mathrm{cm}^6/\mathrm{s}$~\cite{Burt1997}. Using a broken symmetry approach, we write the Bose field operator as a mean field plus fluctuations,
\begin{eqnarray}
\hat{\Psi}(\rbf) = \psi(\rbf) + \hat{\delta}(\rbf) \mathrm{,}
\end{eqnarray}
and substitute this into the previous expression.  Identifying the condensate density as $n_0(\rbf) = |\psi(\rbf)|^2$ yields
\begin{eqnarray}
\dot{n}(\rbf)_\mathrm{3B} &=& -K_3 \Big\{ [n_0(\rbf)]^3 + 9 [n_0(\rbf)]^2 \langle \hat{\delta}^{\dag}(\rbf)\hat{\delta}(\rbf) \rangle \nonumber\\
&&+ 9 n_0(\rbf) \langle \hat{\delta}^{\dag}(\rbf)^2\hat{\delta}(\rbf)^2 \rangle + \langle \hat{\delta}^{\dag}(\rbf)^3\hat{\delta}(\rbf)^3 \rangle \Big\} \mathrm{.}
\end{eqnarray}
The noncondensate density is given by $n_\mathrm{th}(\rbf) = \langle \hat{\delta}^{\dag}(\rbf)\hat{\delta}(\rbf) \rangle$.  Using Wick's theorem on the higher order operator moments of the fluctuations gives
\begin{eqnarray}
\langle \hat{\delta}^{\dag}(\rbf)^2\hat{\delta}(\rbf)^2 \rangle
&=& 2 [n_\mathrm{th}(\rbf)]^2 \mathrm{,} \\
\langle \hat{\delta}^{\dag}(\rbf)^3\hat{\delta}(\rbf)^3 \rangle
&=& 6 [n_\mathrm{th}(\rbf)]^3 \mathrm{,}
\end{eqnarray}
and so
\begin{eqnarray}
\dot{n}(\rbf)_\mathrm{3B} = -K_3 \Big\{ [n_0(\rbf)]^3 + 9 [n_0(\rbf)]^2 n_\mathrm{th}(\rbf)
\nonumber\\
+ 18 n_0(\rbf)[n_\mathrm{th}(\rbf)]^2 +6 [n_\mathrm{th}(\rbf)]^3 \Big\} \mathrm{.}
\label{eq:3bdyloss}
\end{eqnarray}
The loss rates for the condensate and thermal cloud atoms can then be written separately by noting that, for example, the second term in this above expression represents a three-body process in which two condensate atoms and one thermal cloud atom are lost. The coefficients can be divided accordingly to give \cite{Soding1999a}
\begin{eqnarray}
\dot{n}_0(\rbf)_\mathrm{3B} & = & -K_3 \Big\{ [n_0(\rbf)]^3 + 6 [n_0(\rbf)]^2 n_\mathrm{th}(\rbf) 
\nonumber\\
&&+ 6 n_0(\rbf)[n_\mathrm{th}(\rbf)]^2 \Big\} \mathrm{,} \label{eq:3Bc} \\
\dot{n}_\mathrm{th}(\rbf)_\mathrm{3B} & = & -K_3 \Big\{ 3[n_0(\rbf)]^2 n_\mathrm{th}(\rbf) + 12n_0(\rbf)[n_\mathrm{th}(\rbf)]^2 
\nonumber\\
&&+ 6[n_\mathrm{th}(\rbf)]^3 \Big\} \mathrm{,}
\end{eqnarray}
which ensures $\dot{n}_0(\rbf)_\mathrm{3B}+\dot{n}_\mathrm{th}(\rbf)_\mathrm{3B}=\dot{n}(\rbf)_\mathrm{3B}$.
To incorporate these loss rates in our semiclassical theory (Appendix~\ref{sec:sc}), we treat these expressions for $\dot{n}_0(\rbf)_\mathrm{3B}$ and $\dot{n}_\mathrm{th}(\rbf)_\mathrm{3B}$ as coupled ordinary differential equations, and we solve them in time at each spatial gridpoint via Euler's method. This is not entirely straightforward, because the semiclassical theory is strictly static: The final equilibrium is determined from the initial equilibrium in a single leap, without any stepwise time evolution. In the case of sudden dimple turn-on, we begin with the final equilibrium densities, and we calculate three-body loss over the 1-s equilibration time. In essence, we are assuming that the three-body loss rates on the final equilibrium densities are approximately equal to the average loss rates during the 1-s evolution from initial to final equilibrium. In the case of the quasistatic dimple turn-on, we can be more precise: Because the system is never out of equilibrium as it evolves from initial to final equilibrium, we can break up the process into arbitrarily many steps. For each step, we evolve the system isentropically as the dimple depth is incrementally increased, and we calculate three-body loss over the corresponding time interval.

\begin{figure}[t]
  \begin{center}
		\begin{psfrags}%
		\psfragscanon%
		%
		\psfrag{s05}[t][t]{\color[rgb]{0,0,0}\setlength{\tabcolsep}{0pt}\begin{tabular}{c}Dimple Depth 		($k_\mathrm{B}$nK)\end{tabular}}%
		\psfrag{s06}[b][b]{\color[rgb]{0,0,0}\setlength{\tabcolsep}{0pt}\begin{tabular}{c}Condensate Fraction 		(\%)\end{tabular}}%
		\psfrag{s07}[l][l]{\color[rgb]{0,0,0}\setlength{\tabcolsep}{0pt}\begin{tabular}{l}(a)\end{tabular}}%
		\psfrag{s08}[t][t]{\color[rgb]{0,0,0}\setlength{\tabcolsep}{0pt}\begin{tabular}{c}Time since dimple turn-on 			(ms)\end{tabular}}%
		\psfrag{s10}[l][l]{\color[rgb]{0,0,0}\setlength{\tabcolsep}{0pt}\begin{tabular}{l}(b)\end{tabular}}%
		%
		\psfrag{x01}[t][t]{200}%
		\psfrag{x02}[t][t]{400}%
		\psfrag{x03}[t][t]{600}%
		\psfrag{x04}[t][t]{800}%
		\psfrag{x05}[t][t]{1000}%
		\psfrag{x06}[t][t]{0}%
		\psfrag{x07}[t][t]{500}%
		\psfrag{x08}[t][t]{1000}%
		\psfrag{x09}[t][t]{1500}%
		%
		\psfrag{v01}[r][r]{0}%
		\psfrag{v02}[r][r]{5}%
		\psfrag{v03}[r][r]{10}%
		\psfrag{v04}[r][r]{15}%
		\psfrag{v05}[r][r]{20}%
		\psfrag{v06}[r][r]{25}%
		%
		\includegraphics[width=0.47\textwidth]{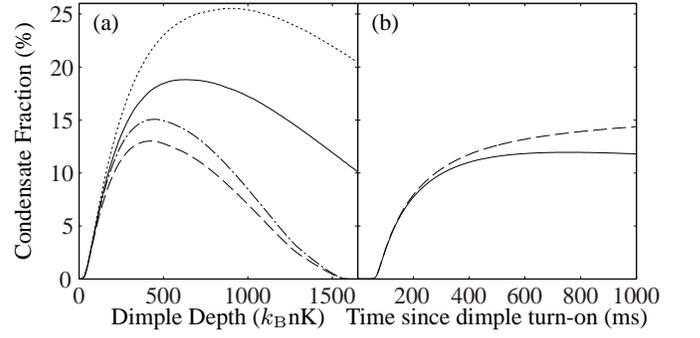}%
		\end{psfrags}%
  \end{center}
\caption{The effects of three-body loss on the predicted condensate fraction. (a) Semiclassical Hartree-Fock predictions of the equilibrium condensate fraction vs dimple depth in a narrow dimple for adiabatic turn-on, including three-body loss (solid line), adiabatic turn-on, with no three-body loss (dotted line), sudden turn-on, including three-body loss (dashed line), and sudden turn-on, with no three-body loss (dot-dash line). (b) Quantum kinetic theory predictions of condensate growth following the sudden turn-on of the narrow dimple of depth $A/\kB=660~nK$, including three-body loss (solid line) and with no there-body loss (dashed line).}
\label{fig:3Bloss}
\end{figure}

The inclusion of three-body loss has a significant effect on the predictions of both semiclassical Hartree-Fock and quantum kinetic theory, as we show in Fig.~\ref{fig:3Bloss}. This is particularly evident in the case of adiabatic turn-on, where the inclusion of three-body loss not only drastically reduces the maximum condensate fraction but also reduces the optimal dimple depth at which the maximum fraction is attained. All of the curves shown correspond to the narrow dimple, with initial conditions $N=2\times 10^5$ and $T_\mathrm{i}=150$~nK.

\section{Ergodic quantum Boltzmann equation}
\label{sec:EQBE}

In our implementation of the EQBE, we use an energy grid with uniform spacing $\delta\epsilon$, with the lowest energy bin corresponding to the condensate. Because the condensate energy level (equal to the bottom of the effective potential) changes with time, we redefine our energy grid to be $\epsilon' = \epsilon - \mu[N_0(t)]$, where $\mu[N_0(t)]$ is the condensate chemical potential in the Thomas-Fermi approximation. With this transformation the EQBE is~\cite{Bijlsma2000c}
\begin{equation} \label{eq:EQBE}
	\dot{n}(\epsilon,t) = -\frac{\partial}{\partial\epsilon}\left[\frac{g_\mathrm{w}(\epsilon,t)}{g(\epsilon,t)}n(\epsilon,t)\right]
												+ I_\mathrm{coll}(\epsilon,t) \mathrm{,}
\end{equation}
where $g_\mathrm{w}(\epsilon,t)$ is the weighted density of states, defined as
\begin{equation}
	g_\mathrm{w}(\epsilon,t)=\int \frac{d\rbf d\pbf}{(2\pi\hbar)^3} 
  						\delta\big(\epsilon-\epsilon_{\mathrm{th}}(\rbf,\pbf,t)\big)\frac{\partial V(\rbf,t)}{\partial t} \mathrm{,}
\end{equation}
which contains the time derivative of the effective potential $V(\rbf,t)=V_\mathrm{ext}(\rbf)+2U_0n_0(\rbf,t)$ and hence depends on the rate of condensate growth $\dot{N}_0(t) = \dot{n}(0,t) = I_\mathrm{coll}(0,t)$. The contribution of binary elastic collisions is given by the term
\begin{eqnarray} \label{eq:Icoll}
	I_\mathrm{coll}(\epsilon,t)=\frac{m^3 g^2}{2\pi^3\hbar^7} \int d\epsilon_2 d\epsilon_3 d\epsilon_4 g(\epsilon_\mathrm{min})
 \delta(\epsilon + \epsilon_2 - \epsilon_3 - \epsilon_4) 
\nonumber\\
\times\left[ (1+f)(1+f_2)f_3 f_4 - ff_2(1+f_3)(1+f_4)\right] \mathrm{,}
\nonumber \\
\end{eqnarray}
where $f=f(\epsilon,t)$, $f_i=f(\epsilon_i,t)$, and $\epsilon_\mathrm{min}=\mathrm{min}\{\epsilon,\epsilon_2,\epsilon_3,\epsilon_4\}$. The first (second) term within the square brackets represents the forward (backward) collisions $\epsilon_3 + \epsilon_4 \leftrightarrow \epsilon + \epsilon_2$. Factors of the form $(1+f_i)$ are due to Bose enhancement and vanish in the classical limit $f_i\ll 1$. For the case of collisions involving a condensate atom, $I_\mathrm{coll}(0,t)$, we make the replacement $g(0)[1+f(0,t)] \approx g(0)f(0,t) = N_0(t)$. Detailed derivations of the EQBE can be found in~\cite{Luiten1996a,Davis2000a,Bijlsma2000c}.

We numerically evolve the EQBE in time using an explicit fourth-order Runge-Kutta method. At each time step, we calculate $I_\mathrm{coll}$ by summing over all combinations of energy bins $\{\epsilon,\epsilon_2,\epsilon_3,\epsilon_4\}$ satisfying the delta function in Eq.~(\ref{eq:Icoll}) and updating $I_\mathrm{coll}(\epsilon_i,t)$ for all four bins involved in each combination. We then add the contributions from three-body loss [see Eq.~(\ref{eq:3B0QB}) and (\ref{eq:3BthQB})], and use the resulting value of $\dot{N}_0(t)$ to calculate $g_\mathrm{w}(\epsilon,t)$ in the first term on the right-hand side of the EQBE.

To incorporate three-body loss in our quantum kinetic calculations, the loss rate for the thermal cloud must be modified to include energy dependence:
\begin{eqnarray} \label{eq:3Bth}
\dot{n}_\mathrm{th}(\epsilon,\rbf)_\mathrm{3B} &=& -K_3 n_\mathrm{th}(\epsilon,\rbf) \Big\{ 3[n_0(\rbf)]^2 
\nonumber\\
&+& 12n_0(\rbf)n_\mathrm{th}(\rbf) + 6[n_\mathrm{th}(\rbf)]^2 \Big\} \mathrm{,}
\end{eqnarray}
where the energy-position density of the thermal cloud is calculated from the energy density as
\begin{equation}
n_\mathrm{th}(\epsilon,\rbf)=g(\epsilon,\rbf)f(\epsilon)=\frac{g(\epsilon,\rbf)}{g(\epsilon)}n(\epsilon)\mathrm{,}
\end{equation}
and $g(\epsilon,\rbf)$ is the position-dependent density of states, obtained by omitting the spatial integral from Eq.~(\ref{eq:dos}). At each time step in the numerical evolution of the EQBE, we determine the condensate density $n_0(\rbf)$ from the Thomas-Fermi approximation [Eq.~(\ref{eq:TF}), with the thermal cloud mean-field neglected] and determine the thermal cloud density $n_\mathrm{th}(\rbf)$ by integrating this expression for $n_\mathrm{th}(\epsilon,\rbf)$ over energy. We then calculate the loss rate for the condensate, Eq.~(\ref{eq:3Bc}), and the energy-dependent loss rate for the thermal cloud, Eq.~(\ref{eq:3Bth}). Lastly, we integrate out the spatial dependence to get
\begin{eqnarray}
	\dot{N}_{0\,\mathrm{3B}} & = & \int d\rbf \dot{n}_0(\rbf)_\mathrm{3B} \mathrm{,} \label{eq:3B0QB} \\
	\dot{n}_\mathrm{th}(\epsilon)_\mathrm{3B} & = & \int d\rbf \dot{n}_\mathrm{th}(\epsilon,\rbf)_\mathrm{3B} \mathrm{,} \label{eq:3BthQB}
\end{eqnarray}
which are simply added to the EQBE calculation of $\dot{n}(\epsilon)$ in Eq.~(\ref{eq:EQBE}).

\bibstyle{apsrev4-1}

%

\end{document}